\documentclass[12pt]{article}

\usepackage{amsfonts}
\usepackage{amsmath}
\usepackage{amscd,amssymb,amsthm}
\usepackage{hyperref}
\usepackage{graphicx}
\usepackage{pifont}
\usepackage{graphicx}
\graphicspath{{figures/}} % Graphics will be here
\usepackage{multirow}

\usepackage{listings}
\usepackage{bookmark}

\newcommand{\be}{\begin{equation}}
\newcommand{\ee}{\end{equation}}
\newcommand{\bea}{\begin{eqnarray}}
\newcommand{\eea}{\end{eqnarray}}
\newcommand{\cmark}{\ding{51}}
\newcommand{\xmark}{\ding{55}}

\def\4{{\sst{(4)}}}

\begin{document}
\pagenumbering{roman}

\begin{titlepage}

\vspace{3.0cm}

\

\

\

%\centerline{\Large \bf p-branes in Chargeless S-brane Backgrounds}
\centerline{\Large \bf Homogeneous Solutions of Minimal Massive 3D Gravity}
\vspace{1.5cm}

\centerline{Jumageldi Charyyev\footnote{jc7518@nyu.edu}}

\noindent
\centerline{Courant Institute of Mathematical Sciences,}\\
\centerline{New York University, NY 10012, USA} 

\

\centerline{Nihat Sadik Deger\footnote{sadik.deger@boun.edu.tr}}

\noindent
\centerline{Department of Mathematics, Bogazici University,}\\
\centerline{Bebek, 34342, Istanbul-Turkey}

\

\centerline{Feza Gursey Center for Physics and Mathematics,}
\centerline{Bogazici University, Kandilli, 34684, Istanbul-Turkey}

\

\vspace{1.5cm}

\centerline{\bf ABSTRACT}
\vspace{0.5cm}
In this paper we systematically construct simply transitive homogeneous spacetime solutions of the three-dimensional Minimal Massive 
Gravity (MMG) model. In addition to those that have analogs in Topologically Massive Gravity, such as warped AdS and pp-waves, there are 
several solutions genuine to MMG. Among them, there is a stationary Lifshitz metric with the dynamical exponent $z=-1$ and an anisotropic 
Lifshitz solution where all coordinates scale differently. Moreover, we identify 
a homogeneous Kundt type solution at the chiral point of the theory. We also show that in a particular limit of the physical parameters in
which the Cotton tensor drops out from the MMG field equation, homogeneous solutions exist only at the merger point in the parameter space 
if they are not conformally flat.

\vspace{2cm}

\end{titlepage}

\pagenumbering{arabic}

\tableofcontents

\section{Introduction}

Minimal Massive Gravity (MMG) is a pure 3-dimensional gravity model proposed in \cite{[Berg]}, which attracted much attention during the last three years.
It is an extension of another widely studied theory known as `Topologically Massive Gravity' (TMG) \cite{[Des]} with a particular curvature squared term in 
the field equation.  It is `minimal' in the sense that there is only one propagating spin-2 mode in the bulk like TMG. However, unlike TMG, it avoids the 
bulk-boundary unitarity clash since in a certain range of its parameters it is possible to have central charge of the dual CFT and energy of the bulk graviton 
positive simultaneously \cite{[Berg], unit1} (however, see \cite{unit2}) which makes it a potentially useful toy model for understanding quantum gravity in 
4-dimensions.

In this paper we make a systematic investigation of homogeneous spacetime solutions of MMG with Lorentzian signature
and obtain a large number of new ones. We will focus on simply transitive Lie groups where any two points can be related by 
an isometry and stability (isotropy) group of any point is trivial. Since a homogeneous (pseudo)Riemannian manifold $M$ has the form 
of a quotient $G/H$, where $G$ is its group of isometries, which is a Lie group, and H is a closed subgroup of G, 
this means that we take $H$ to be just the identity. In this case $M$ and $G$ can be identified and considering left action of such a Lie group 
on itself one can construct its left-invariant metric up to automorphisms using left-invariant 1-forms. This results in a metric with constant 
coefficients and all curvature calculations become algebraic. In three dimensions, classification of Lie algebras was done by 
Bianchi \cite{[Bia]} and one can systematically check whether  
corresponding metrics are solutions of a particular 3-dimensional model. This method was successfully applied to TMG with vanishing cosmological 
constant in \cite{nutku} and \cite{Ortiz}, and more recently for non-zero cosmological constant in \cite{[Mou]}. This was 
also carried out in \cite{[Bak]}-\cite{[Sia]} (see also \cite{aliev2, aliev3}) for another extension of TMG called `New (or General) Massive Gravity' 
(NMG) \cite{new}. 

Since MMG is closely related to the (cosmological) TMG we will follow analysis in \cite{[Mou]} closely which will make identification of most of the solutions we 
obtain straightforward. In \cite{bayram} it was shown that solutions of TMG which have Segre-Petrov-types N and D are also solutions of MMG 
after a redefinition of parameters. We find that, as should be expected, MMG inherits such homogeneous solutions from TMG which include
warped (A)dS and $(A)dS_2 \times S^1$ solutions obtained in \cite{[Arv]} 
and pp-wave spacetimes \cite{iran}. Some of the remaining solutions turn out to be solutions of TMG as well 
only if the cosmological constant is zero which implies that their scalar curvatures vanish, which is not the case in MMG. Finally, we show 
that there are several homogeneous solutions that are genuine to MMG 
one of which is a stationary Lifshitz spacetime (\ref{lifshitz}) with a dynamical exponent $z=-1$ and an anisotropic Lifshitz solution (\ref{generalizedLifshitz}) where all coordinates scale differently.

In \cite{coley} it was proven that 
3-dimensional constant scalar invariant (CSI) Lorentzian spacetimes are locally either homogeneous or Kundt. For MMG the latter is
studied in \cite{[Deg]}. Thus, the current work fills an important gap in the construction of all CSI solutions of MMG.
We also show that two of the Kundt solutions found in \cite{[Deg]} are also homogeneous and one of them appears at the so-called chiral 
point of the theory that has no TMG limit. 

It is possible to modify MMG field equation by dropping the Cotton tensor \cite{iran, [chile]}. Remarkably, for this case
we find that homogeneous solutions exist only at a particular point in the parameter space called the merger point, if 
they are not conformally flat.

The paper is organized as follows. In the next section we introduce the MMG model and explain our method in more detail. Then in subsequent sections, we go through
all possible homogeneous metrics of 3-dimensional Lie algebras one by one. For each solution, 
we provide a coordinate representation of its metric and in most of the cases are able to identify the corresponding spacetime. In 
section \ref{summary} we summarize our results in Table \ref{table} that includes Segre-Petrov types as was proposed in \cite{pope}, 
make a comparison with TMG and indicate some future directions.

\section{Minimal Massive Gravity}

\indent In this section we will give a brief introduction to MMG 
\cite{[Berg]} and describe our method for constructing its homogeneous solutions.
The theory is defined by the field equation 
\begin{equation}
\label{MMG}
G_{\mu\nu}+ag_{\mu\nu}+bC_{\mu\nu}+cJ_{\mu\nu}=0 \, ,         
\end{equation}
where $G_{\mu\nu}$ is the Einstein tensor and the Cotton tensor $C_{\mu\nu}$, which is symmetric, traceless and covariantly conserved, 
is related to the Schouten tensor $S_{\sigma\nu}$ as
\begin{equation}
\label{cotton}
C^{\mu}_{\;\;\nu}\equiv \frac{1}{\sqrt{-g}}\varepsilon^{\mu\rho\sigma} \nabla_{\rho}S_{\sigma\nu}, \indent 
S_{\sigma\nu} \equiv R_{\sigma\nu}-\frac{1}{4}Rg_{\sigma\nu} \, ,
\end{equation}
with $\varepsilon_{012}=+1$. The $J$-tensor is given as 
\begin{eqnarray}
J^{\mu\nu} \equiv R^{\mu\rho}R^{\nu}_{\;\;\rho}-\frac{3}{4}R^{\mu\nu}R-\frac{1}{2}g^{\mu\nu}(R^{\rho\sigma}R_{\rho\sigma}-\frac{5}{8}R^2) \, .
\end{eqnarray}
It is not covariantly conserved, but instead one finds \cite{[Berg]}:
\begin{equation}
 \sqrt{-g} \nabla_{\mu}J^{\mu\nu} = \varepsilon^{\nu\rho\sigma}S_{\rho}^{\ \tau}C_{\sigma \tau} \, ,
\label{divergence}
 \end{equation}
which is not automatically zero. It follows that the MMG field equation 
(\ref{MMG}) cannot be derived from an action that contains only the metric field \cite{[Berg]}. However, for any solution of the field equation
(\ref{MMG}) one can show that the right hand side of (\ref{divergence}) vanishes which establishes the consistency of the model in a novel way.
Moreover, it is still possible to couple matter \cite{[Arv]} and calculate charges of its solutions \cite{charge}.

Finally, the coefficients $a$, $b$ and $c$ in terms of physical parameters are 
\begin{equation}
a= \frac{\bar{\Lambda}_{0}}{\bar{\sigma}}, \indent  b= \frac{1}{\mu \bar{\sigma}},  \indent c= \frac{\gamma}{\mu^2 \bar{\sigma}}.
\label{parameters}
\end{equation}
When $\gamma=0$ (i.e., $c=0$) the model reduces to the (cosmological) TMG model \cite{[Des]}, where such solutions 
were studied before \cite{nutku} -\cite{[Sia]}.

There are two special points in the parameter space of the MMG theory \cite{[Berg]}. The first is called the 'chiral point' at  which one of the 
central charges vanish 
and is given by:
\begin{equation}
\label{chiral}
\bar{\sigma}+\frac{\gamma}{2} (\bar{\sigma}^2-\frac{\gamma \bar{\Lambda}_{0}}{\mu^2}) \pm \sqrt{\bar{\sigma}^2-\frac{\gamma \bar{\Lambda}_{0}}{\mu^2}}=0 \indent \mathrm{or} 
\indent 1+\frac{c}{2b^2}(1-ac)\pm \sqrt{1-ac}=0.  
\end{equation}
The second one is called the 'merger point' where two possible values of the cosmological constant coincide:
\begin{equation}
\label{merger}
\bar{\Lambda}_{0}=\frac{\mu^2 \bar{\sigma}^2}{\gamma} \indent \mathrm{or} \indent ac=1.
\end{equation}

\indent In order to find homogeneous solutions of MMG we will follow the 
method of \cite{[Mou]} that was successfully used for the (cosmological) TMG \cite{[Des]} model which can be summarized as follows: 
First a Lie algebra basis is fixed for each three-dimensional Lie algebra $\mathfrak{g}$ which induces left-invariant Maurer-Cartan 1-forms. 
A left-invariant metric for the Lie group at the identity is identified by a non-degenerate metric on the Lie algebra up to automorphism group of this 
Lie algebra. Starting from an arbitrary left-invariant metric on the algebra, it is put into a simple form using automorphisms. The metric is expressed in terms of 
left-invariant 1-forms with constant coefficients which implies that all curvature calculations, and hence the MMG field equation (\ref{MMG}), become algebraic. 
This method is different but equivalent to the one used in
\cite{nutku, Ortiz, [Bak], [Sia]} where instead of fixing the Lie algebra basis, an orthonormal frame is chosen \cite{milnor}. Then, $SO(1,2)$ Lorentz transformations are 
used to simplify the structure constants.
We prefer 
the strategy of \cite{[Mou]} since it enables us to compare our solutions with those of (cosmological) TMG \cite{[Des]} directly. Moreover, geometric 
identification of common solutions become trivial.

\indent Instead of solving algebraic equations for the constants in the metric $\{u,v,w,...\}$ in terms of the parameters $\{a,b,c\}$ of the MMG theory
(\ref{parameters}) it is more convenient and illuminating to display the parameters in the theory in terms of the parameters of the metric. This reduces 
to solving a system of linear equations
\begin{equation}
\label{system}
A\cdot \begin{pmatrix}
  a  \\
	  b  \\
		c  \\
		  \end{pmatrix}=V
\end{equation}
for  $\{a,b,c\}$  where $A$ is a matrix is of the dimension $k\times3$ and $V$ is a $k\times 1$ vector with $k=3,4$ or $6$. The number $k$ is 
determined by the number of independent components of the field equation (\ref{MMG}). The rank of the matrix $A$ can 
be at most three. When the rank of $A$ is three, the linear equation (\ref{system}) has a unique solution, provided that the solution exists. If the solution exists, the 
cases when the rank of $A$ is less than three should be considered separately as in such situations new solutions may arise. If $A$ is of 
the dimension $3\times3$, then computing the determinant of $A$ is enough to determine when the rank of $A$ is less than three. When $A$ is not a square 
matrix and for the cases when (\ref{system}) does not have a general solution, a more careful analysis is required. For example, it may happen that for a particular 
relation among the parameters of the metric the system becomes consistent.

\indent To identify distinct Lie algebras one must determine sets of structure constants which cannot be related by linear 
transformations. For three-dimensional Lie algebras this classification was done by Bianchi \cite{[Bia]} but usually presented 
in a more modern approach described in \cite{[Mac]} (see also \cite{[Ste]}.) 
Besides the abelian $\mathbb{R}^3$ and the two familiar algebras $\mathfrak{sl}_2$ and $\mathfrak{su}_2$, we also have the Lie algebras $\mathfrak{a}_\infty$ and 
$\mathfrak{a}_0$, and two continuous families of Lie algebras: $\mathfrak{iso}(1,1;\theta)$ and $\mathfrak{iso}(2;\theta)$ where parameter $\theta$ varies 
in $(0,\frac{\pi}{2}]$. In the first, $\theta$ values $\{0, \frac{\pi}{4}\}$ are special and should be considered separately which in total leads to 9 Bianchi classes.
%\footnote{If $\theta=\frac{\pi}{2}$ cases are also treated separately, then there are 11 classes.} 
Finally, $\mathfrak{iso}(1,1;0)$ and $\mathfrak{iso}(2;0)$ are isomorphic to each other.

\indent We now begin constructing the homogeneous spacetime solutions of MMG going through the above list of algebras. 
We assume that metrics are Lorentzian with mostly plus signature and follow conventions and terminology of \cite{[Mou]} to which we refer for details.

\section{Solutions on \texorpdfstring{$SL(2,\mathbb{R})$}-}

\indent For the Lie algebra $\mathfrak{sl}_2$ of $SL(2,\mathbb{R})$ a basis $\{\tau_0,\tau_1,\tau_2\}$ can be fixed with
\begin{equation}
[\tau_0,\tau_1]=\tau_2, \indent \indent [\tau_2,\tau_1]=\tau_0, \indent \indent [\tau_2,\tau_0]=\tau_1.
\end{equation}
Let $\theta^{a}$ be a dual basis of $\tau_a$. 
Elements of $SL(2,\mathbb{R})$ can be parametrized by a group representative as (see for example \cite{[Mou2]})  
\begin{equation}
\mathcal{V}(x)=e^{t(\tau_0+\tau_2)}e^{\sigma \tau_1}e^{\zeta \tau_2}.
\end{equation}
It follows that the Maurer-Cartan one-forms are 
\begin{equation}
\mathcal{V}^{-1}d\mathcal{V}=(e^\sigma \cosh\zeta dt - \sinh\zeta d\sigma)\tau_0+(\cosh\zeta d\sigma - e^\sigma \sinh\zeta dt)\tau_1+(d\zeta+e^\sigma dt)\tau_2.
\end{equation}
There are 4 classes of left-invariant metrics on $SL(2,\mathbb{R})$ that are given below (see \cite{[Mou]}).

\subsection{111-type metric}

\indent 111-type metric is of the form 
\begin{equation}
g=u\theta^0 \theta^0+v\theta^1 \theta^1+w\theta^2 \theta^2 \, ,
\label{111}
\end{equation}
where $uw<0$ and $v>0$ for Lorentzian, mostly plus signature. The case $-u=v=w$ corresponds to the round AdS$_3$ written as Hopf fibration over 
AdS$_2$ spacetime in the Poincar\'e coordinates. The general line element is 
\begin{eqnarray} \nonumber
 ds^2 & = & e^{2\sigma}(u\cosh^2\zeta + v\sinh^2\zeta)dt^2 + (u\sinh^2\zeta + v\cosh^2\zeta)d\sigma^2 \\ 
 &-& 2(u+v)e^{\sigma}\cosh\zeta \sinh\zeta dt d\sigma + w(d\zeta + e^{\sigma}dt)^2 \, .
\label{111-metric}
\end{eqnarray}

\indent The coefficients $a$, $b$, $c$ and the scalar curvature $R$ in terms of $u$, $v$, and $w$ are 
\begin{eqnarray} \nonumber 
a&=&\frac{1}{Q}\cdot \frac{[(u+v+w)^2-4vw]^3}{8uvw} \, , \\ \nonumber
b&=&-\frac{1}{Q}\cdot 8\sqrt{-uvw}[u^2-(v-w)^2](u+v+w) \, , \\ \nonumber
c&=&\frac{1}{Q}\cdot 8uvw[(u+v+w)^2-4vw] \, , \\ 
R&=&-\frac{(u+v+w)^2-4vw}{2uvw} = - \frac{c\, Q}{16(uvw)^2} = - \frac{(a \, Q)^{1/3}}{(uvw)^{2/3}} \,\, ,
\label{111-general}
\end{eqnarray}
where
\begin{equation}
Q= [(u+v+w)^2-4vw]^2+8[u^2-(v+w)^2][u^2-(v-w)^2] \, .
\end{equation}
This solution in general represents a triaxially deformed AdS spacetime. Note that when $c=0$, i.e. for TMG, $R=0$. 
In this case, the cosmological constant $a$ vanishes too. Also, it can be shown that when $R=0$ we have $Q \neq 0$. Therefore, for this 
solution $Q$ is non-zero, since $Q$ and $R$ cannot vanish at the same time.

The matrix $A$ in the equation (\ref{system}) is a $3\times3$ matrix with the determinant
\begin{equation}
detA=-Q\cdot \frac{(u+v)(v-w)(u+w)}{8(-uvw)^{5/2}}.
\end{equation}
Thus the cases $Q=0$, $u=-v$ (or equivalently $u=-w$) and $v=w$ should be considered separately. We have checked that the case $Q=0$ does not give rise to
any solution to (\ref{MMG}).

\indent 
\pmb{$u=-v$}: In this case the spacetime metric (\ref{111-metric}) becomes:
\begin{equation}
\label{spacelike}
ds^2= v[-e^{2\sigma}dt^2 +d\sigma^2] + w(d\zeta+e^{\sigma}dt)^2 \equiv g_{(2)} + w(d\zeta+ \chi)^2 \, ,
\end{equation}
where $w>0$. This solution was found before in \cite{[Arv]} and is called the spacelike warped\footnote{Constants $v$ and $w$ are related 
to the warping parameter $\nu$ in \cite{[Arv]} as $v=\frac{l^2}{(\nu^2+3)}, \indent w=\frac{4l^2\nu^2}{(\nu^2+3)^2}$. The 
limit $\nu\rightarrow 1$ in (\ref{spacelike}) corresponds to the AdS metric, where $-u=v=w$. }
AdS. Note that $d\chi$=vol$ g_{(2)}$. The coefficients $a$ and $b$ in terms of $u$, $v$, and 
$w$ are 
\begin{equation}
a=\frac{16v^2(w-4v)+c(4v-7w)(4v-3w)}{192v^4}, \indent b=\frac{8v^2+c(4v-3w)}{12\sqrt{v^2w}}.
\label{111-spacelike}
\end{equation}
The curvature scalar is as in (\ref{111-general}).

\indent 
\pmb{$v=$}\pmb{$w$}: In this case $u<0$ and the spacetime metric (\ref{111-metric}) becomes:
\begin{equation}
ds^2=u(e^\sigma \cosh\zeta dt - \sinh\zeta d\sigma)^2+w[(\cosh\zeta d\sigma - e^\sigma \sinh\zeta dt)^2+ (d\zeta+e^\sigma dt)^2] \, .
\label{timelike}
\end{equation}
This metric was identified as timelike warped AdS in \cite{[Mou]} and the coefficients $a$ and $b$ are:
\begin{equation}
a=\frac{-16w^2(u+4w)+c(3u+4w)(7u+4w)}{192w^4}, \indent b=\frac{8w^2+c(3u+4w)}{12w \sqrt{-u}}.
\label{111-timelike}
\end{equation}
Again the curvature scalar is given in (\ref{111-general}). 

%It is worthwhile to note that in all the above cases when $b=0$ we are at the merger point (\ref{merger}).

\subsection{12-type metric}

\indent 12-type metric is of the form 
\begin{equation}
g=v(-\theta^0\theta^0+\theta^1\theta^1)+w\theta^2\theta^2+z(\theta^0+\theta^1)^2 \, ,
\label{12}
\end{equation}
with $z\neq0$, $v>0$ and $w>0$. Here, $z$ can be scaled to $\pm 1$. Notice that, it 
is a $z$-deformation of the spacelike warped AdS metric (\ref{spacelike}).

\indent The coefficients $a$, $b$, $c$ and the scalar curvature $R$ in terms of $v$ and $w$ are 
\begin{eqnarray} \nonumber
a&=& \frac{1}{Q}\cdot\frac{(w-4v)^3}{8v^2} \, , \\ \nonumber
b&=& \frac{1}{Q}\cdot8(2v-w)\sqrt{v^2w} \, , \\   \nonumber
c&=& \frac{1}{Q}\cdot8v^2(w-4v) \, , \\ 
R&=& \frac{w-4v}{2v^2} \, ,
\label{12-general}
\end{eqnarray}
where $Q=-(w-4v)^2+8(4v^2-w^2)$. Note that in the TMG limit, i.e., $c=0$, both the scalar curvature and cosmological constant $a$ vanish.
Adapting the coordinate transformations given in \cite{[Mou]} to our case as:
\begin{equation}
 t= \frac{1}{2x} +\frac{y}{2l^2} \, , \indent \, e^{\sigma} =2x \, , \indent \, \zeta = \frac{\rho}{kl} +\ln x \, ,
\label{transformation}
 \end{equation}
we obtain
\begin{equation}
 ds^2= d\rho^2 + 2dydx + (\frac{R}{2} + \frac{3k^2}{4l^2})x^2 dy^2 + \frac{2k}{l} x d\rho dy + \frac{z}{l^4} e^{-\frac{2}{kl}\rho} dy^2\, ,
\label{12-metric}
 \end{equation}
where $v=l^2$ and $w=k^2l^2$ with $k > 0$. This solution is Kundt type and corresponds to a special case found 
in \cite{[Deg]}, namely its equation (58) with some particular choices.

$A$ in equation (\ref{system}) is a $4\times3$ matrix. Only $Q=0$ and $v=w$ cases should be considered separately 
and the first does not provide any solution.

\pmb{$v=$}\pmb{$w$}: Here the coefficients $a$ and $b$ are equal to
\begin{equation}
a=-\frac{16w+c}{64w^2}, \indent b=\frac{8w+c}{12\sqrt{w}}.
\label{12-null warped}
\end{equation}
In this case the coefficient of the third term in the metric (\ref{12-metric}) above vanishes since $k=1$. By
defining a new  coordinate $\theta= x e^{\rho/l}$ it becomes
\begin{equation}
 ds^2= d\rho^2 + 2 e^{-\frac{\rho}{l}} dy d\theta + \frac{z}{l^4} e^{-\frac{2\rho}{l}} dy^2\, ,
 \label{12-null warped metric}
\end{equation}
which corresponds to the null warped AdS (Schr\"odinger) spacetime that was obtained before in \cite{[Arv]}.

%Note that in all the above cases when $b=0$ we get the merger point (\ref{merger}).

\subsection{3-type metric}

\indent 3-type metric is of the form 
\begin{equation}
g=v(-\theta^0\theta^0+\theta^1\theta^1+\theta^2\theta^2) + z(\theta^0\theta^2+\theta^1\theta^2) \, ,
\end{equation}
with $z\neq0$ and $v>0$. Note that it is a $z$-deformation of the metric (\ref{111}) 
with $-u=v=w$ (i.e., round AdS). The constant $z$ can be scaled to $\pm 1$.

\indent The coefficients $a$, $b$, $c$ and the scalar curvature $R$ in terms of $v$ are equal to
\begin{equation}
\label{3-type}
a=-\frac{9}{40v}, \indent b=\frac{8\sqrt{v}}{15}, \indent c=-\frac{8v}{5}, \indent R=-\frac{3}{2v}.
\end{equation}
Note that the $z$-deformation has no affect on the scalar curvature which is the same as round AdS$_3$.  
Moreover, the solution is attained at the chiral point, i.e., the coefficients satisfy the equality (\ref{chiral}) with the plus sign. 

Using the coordinate transformations given above (\ref{transformation}) with $k=1$ we obtain
\begin{equation}
 ds^2 = d\rho^2 + 2dydx + (\frac{2x}{l} + \frac{z}{l^3}e^{-\rho/l}) dy d\rho +\frac{z}{l^4} e^{-\rho/l} x dy^2 \, ,
 \label{3-type metric}
\end{equation}
where $v=l^2$. This is a particular case of a Kundt solution given in equation (47) of \cite{[Deg]}.

\subsection{\texorpdfstring{$1z\bar{z}-$}-type metric}

\indent $1z\bar{z}$-type metric is of the form 
\begin{equation}
g=v(-\theta^0\theta^0+\theta^1\theta^1)+w\theta^2\theta^2+2z\theta^0\theta^1 \, ,
\end{equation}
with $vz\neq0$ and $w>0$. Like (\ref{12}) it is a deformation of the spacelike warped AdS metric (\ref{spacelike}). 
When $v=w$, then this solution is a deformation of round AdS. The line element is
\begin{equation}
 ds^2= [v+ z\sinh2\zeta](-e^{2\sigma}dt^2 +d\sigma^2) + w(d\zeta+e^{\sigma}dt)^2 +2ze^{\sigma}\cosh2\zeta d\sigma dt -2z\sinh2\zeta d\sigma^2 
\label{1zz-metric}
 \end{equation}
which was identified with type b) solution of \cite{Ortiz} in \cite{[Mou]}.

The coefficients $a$, $b$, $c$ and the scalar curvature $R$ are equal to
\begin{eqnarray} \nonumber
a&=&\frac{1}{Q}\cdot \frac{(4vw-w^2+4z^2)^3}{8w(v^2+z^2)} \, , \\ \nonumber
b&=&\frac{1}{Q}\cdot 8(w-2v)(w^2+4z^2)\sqrt{w(v^2+z^2)} \, , \\ \nonumber
c&=&\frac{1}{Q}\cdot 8w(4vw-w^2+4z^2)(v^2+z^2) \, , \\ 
R&=&-\frac{4vw-w^2+4z^2}{2w(v^2+z^2)} \, ,
\label{1zz-general}
\end{eqnarray}
where $Q=(4vw-w^2+4z^2)^2 +8(w^2-4v^2)(w^2+4z^2)$. Notice that unlike TMG for which $c=a=0$, in MMG the scalar curvature can be non-vanishing.

Here the matrix $A$ defined in (\ref{system}) is a $4\times3$ matrix. The only special case that should 
be considered separately is when $Q=0$, which does not produce any solution to (\ref{MMG}). 

%We also note that when $b=0$ the merger point equality (\ref{merger}) holds.

\section{Solutions on \texorpdfstring{$SU(2)$}-}

\indent We fix a basis $\{\tau_1,\tau_2,\tau_3\}$ and its dual basis $\theta^{a}$ for the Lie algebra $\mathfrak{su}_2$ with
\begin{equation}
[\tau_1,\tau_2]=\tau_3, \indent \indent [\tau_2,\tau_3]=\tau_1, \indent \indent [\tau_3,\tau_1]=\tau_2.
\end{equation}
An element of $SU(2)$ can be parametrized by (see \cite{[Mou2]})
\begin{equation}
\mathcal{V}=e^{\phi \tau_3}e^{\xi \tau_2}e^{\psi \tau_3}.
\end{equation}
The Maurer-Cartan one-forms are
\begin{eqnarray} \nonumber
\mathcal{V}^{-1}d\mathcal{V}&=&(\sin\psi d\xi -\cos\psi \sin\xi d\phi )\tau_1 \\ 
&+&(\cos\psi d\xi +\sin\psi \sin\xi d\phi )\tau_2+(d\psi+\cos\xi d\phi )\tau_3.
\end{eqnarray}
A left-invariant metric $g$ on $SU(2)$ can be written as: 
\begin{equation}
g=u\theta^1 \theta^1+v\theta^2 \theta^2+w\theta^3 \theta^3 \, ,
\end{equation}
with $uvw < 0$, not all negative.
The coefficients $a$, $b$, $c$ and the scalar curvature $R$ are
\begin{eqnarray} \nonumber
a&=&\frac{1}{Q}\cdot \frac{[(u-v-w)^2-4vw]^3}{8uvw} \, , \\ \nonumber
b&=&\frac{1}{Q}\cdot 8\sqrt{-uvw} [u^2-(v-w)^2](u-v-w) \, ,  \\ \nonumber
c&=&\frac{1}{Q}\cdot 8uvw[(u-v-w)^2-4vw]\, ,  \\
R&=&-\frac{(u-v-w)^2-4vw}{2uvw} \, ,
\label{su2-general}
\end{eqnarray}
where 
\begin{equation} 
Q= [(u-v-w)^2-4vw]^2+8[u^2-(v+w)^2][u^2-(v-w)^2] \, .
\end{equation}
The line element is given by
\begin{equation}
ds^2=(u-v)(\sin\psi d\xi -\cos\psi \sin\xi d\phi )^2+v(d\xi^2+\sin^2\xi d\phi^2)+w(d\psi+\cos\xi d\phi)^2 \, ,
\label{su2-general metric}
\end{equation}
which corresponds to a triaxially deformed sphere and in MMG the scalar curvature given in (\ref{su2-general}) is non-vanishing, unlike TMG.

Moreover, $A$ in equation (\ref{system}) is a $3\times3$ matrix with the determinant
\begin{equation}
detA=\frac{Q}{8(uvw)^3}\cdot \sqrt{-uvw}(u-v)(u-w)(v-w).
\end{equation}
Thus, the cases $Q=0$ and $u=v$ (is enough due to the symmetry) should be considered separately. Again the case $Q=0$ does not give any solution.

\pmb{$u=v$}:  Note that in this case $w < 0$. The coefficients $a$ and $b$ are
\begin{equation}
a=\frac{16v^2(4v-w)+c(4v-3w)(4v-7w)}{192v^4}, \indent b=-\frac{8v^2+c(3w-4v)}{12\sqrt{-v^2w}}.
\label{squashed}
\end{equation}
In this case the metric (\ref{su2-general metric}) simplifies to
a Hopf-fibration over $S^2$. Depending on whether $|w|> 1$ or $|w|< 1$, we have stretched or squashed warpings respectively.

%In all the above cases when $b=0$ the merger point condition (\ref{merger}) is fulfilled.

\section{Solutions on \texorpdfstring{$A_{\infty}$}-}

\indent The Lie algebra $\mathfrak{a}_\infty$ of $A_{\infty}$ is spanned by $r$, $x$, and $y$ and has only one non-trivial bracket 
\begin{equation}
 [r,x]=-y \, .
\end{equation}
We denote the dual basis as $\{\tilde{r}, \tilde{x}, \tilde{y}\}$. The Baker-Campbell-Hausdorff formula allows us to write a representative as (see \cite{[Mou2]})
\begin{equation}
\mathcal{V}=e^{sr}e^{tx}e^{\rho y}.
\end{equation}
The Maurer-Cartan one-forms are
\begin{equation}
\mathcal{V}^{-1}d\mathcal{V}=(ds)r+(dt)x+(d\rho-tds)y.
\end{equation}
By the automorphism group, a left-invariant metric can be fixed as \cite{[Mou]} 
\begin{equation}
g=u\tilde{r}\tilde{r}+v\tilde{x}\tilde{x}\pm \tilde{y}\tilde{y}.
\end{equation}
where $uv \neq 0$ and $u$ or $v$ can be scaled to $\pm 1$. The line element reads as
\begin{equation}
ds^2=uds^2+vdt^2 \pm (d\rho-tds)^2 \, ,
\label{ainf metric}
\end{equation}
which is a Hopf fibration over a flat space. We have
\begin{eqnarray} \nonumber
a=\frac{16|uv|+21c}{192(uv)^2} \, \, , \, \, 
b=\pm\frac{8|uv|- 3c}{12\sqrt{|uv|}} \, \, , \, \,  
R=\frac{1}{2|uv|}.
\label{ainf}
\end{eqnarray}

%Notice that when $b=0$ we are at the merger point (\ref{merger}). 

\section{Solutions on \texorpdfstring{$A_0$}-}

\indent The Lie algebra $\mathfrak{a}_0$ of $A_0$, spanned by $r$, $x$, and $y$, has non-vanishing brackets
\begin{equation}
[r,x]=x, \indent \indent [r,y]=x+y.
\end{equation}
We denote the dual basis as $\{\tilde{r}, \tilde{x}, \tilde{y}\}$. Again by the Baker-Campbell-Hausdorff formula we can choose the representative
\begin{equation}
\mathcal{V}=e^{\xi x+\rho y}e^{\alpha r}.
\end{equation}
Then the Maurer-Cartan one-forms are
\begin{equation}
\mathcal{V}^{-1}d\mathcal{V}=(e^{-\alpha}d\xi -\alpha e^{-\alpha}d\rho)x+(e^{-\alpha}d\rho )y+(d\alpha)r.
\end{equation}
The following 4 types of metrics are available \cite{[Mou]}:

\subsection{\texorpdfstring{$B_1$-}-type metric} \label{a0}

\indent Metric is given by 
\begin{equation}
B_1=z\tilde{r}^2 \pm \tilde{x}^2+v\tilde{y}^2.
\label{a0b1}
\end{equation}
There is no solution for $a$, $b$, $c$.

\subsection{\texorpdfstring{$B_2$-}-type metric}

\indent Metric is given by 
\begin{equation}
B_2=z\tilde{r}^2\pm2\tilde{x}\tilde{y} \, ,
\label{b2}
\end{equation}
with $z>0$. The MMG field equation (\ref{MMG}) is satisfied if 
\begin{equation}
\label{ch2}
a=-\frac{4z+c}{4z^2}, \indent b=\mp\frac{2z+c}{2\sqrt{z}}, \indent R=-\frac{6}{z}.
\end{equation}
Note that the solution is attained at the chiral point (\ref{chiral}). 

%Moreover, when $b=0$ we have the merger point (\ref{merger}). 

Under the coordinate transformations
\begin{equation}
\alpha \rightarrow \log(w), \indent \rho\rightarrow -lx^+, \indent \xi\rightarrow lx^-
\end{equation}
where $z=l^2$ the metric (\ref{b2}) becomes
\begin{equation}
ds^2= \mp \frac{l^2}{w^2}[2\log(w)(dx^+)^2+2dx^+dx^-\mp dw^2]
\label{ch2 metric}
\end{equation}
which is the logarithmic pp-wave solution found in \cite{iran}.

\subsection{\texorpdfstring{$B_3$-}-type metric}

\indent Metric is given by 
\begin{equation}
B_3=z\tilde{r}^2+\tilde{r}\tilde{x}+v\tilde{y}^2.
\end{equation}
In this case $a=0$ is the necessary and sufficient condition to solve the equation (\ref{MMG}). The Ricci, Cotton, and $J$-tensors are identically zero. 
Metric of this Ricci flat spacetime is
\begin{equation}
ds^2=zd\alpha^2+e^{-\alpha}(d\xi d\alpha -\alpha d\rho d\alpha)+ve^{-2\alpha}d\rho^2 \, .
\label{cot1}
\end{equation}
As we discuss in section \ref{summary}, it must be maximally symmetric based on a result of \cite{alex} and hence should locally be Minkowski spacetime.

\subsection{\texorpdfstring{$B_4$-}-type metric}

\indent Metric is given by 
\begin{equation}
B_4=z\tilde{r}^2+\tilde{r}\tilde{y}+u\tilde{x}^2 \, ,
\end{equation}
where $u>0$. The coefficients $a$, $b$, $c$ and the scalar curvature $R$ are found to be
\begin{equation}
\label{1}
a=-\frac{u}{18}, \indent b=-\frac{4}{9\sqrt{u}}, \indent c=-\frac{2}{9u}, \indent R=2u.
\end{equation}
The line element is
\begin{equation}
 ds^2= z d\alpha^2 + e^{-\alpha} d\rho d\alpha + u e^{-2\alpha} (d\xi - \alpha d\rho)^2
 \label{1 metric}
\end{equation}
Unfortunately, we could not determine which spacetime geometry this metric corresponds.
Higher order curvature scalars are as follows
\begin{equation} 
R_{\mu \nu}R^{\mu \nu}=12u^2 \, \, , \, \,
R_{\mu \rho}R^{\rho \nu}R_{\nu}^{\ \mu}=8u^3 \, .
\end{equation}

\section{Solutions on \texorpdfstring{$ISO(2;\theta)$}-}

\indent Let the Lie algebra basis and the dual basis of $\mathfrak{iso}(2;\theta)$ be $\{l, m_1, m_2\}$ and 
$\{\tilde{l}, \tilde{m_1}, \tilde{m_2}\}$ respectively. The non-vanishing brackets are
\begin{equation}
[l,m_1]=2\cos\theta m_1+2\sin\theta m_2, \indent [l,m_2]=2\cos\theta m_2-2\sin\theta m_1 \, ,
\end{equation}
where $\theta \in [0,\pi/2]$. 
We choose the group representative
\begin{equation}
\mathcal{V}=e^{xm_1+ym_2}e^{\rho l}.
\end{equation}
Then the Maurer-Cartan one-forms are
\begin{eqnarray} \nonumber
\mathcal{V}^{-1}d\mathcal{V}=(d\rho)l&+&e^{-2\rho \cos\theta}[\cos(2\rho \sin\theta)dx+\sin(2\rho \sin\theta)dy]m_1 \\
&+&e^{-2\rho \cos\theta}[-\sin(2\rho \sin\theta)dx+\cos(2\rho \sin\theta)dy]m_2.
\end{eqnarray}
There are two types of metrics as below \cite{[Mou]}. The case $\theta=0$ should be analyzed separately.

\subsection{\texorpdfstring{$B_1$-}-type metric}
\label{secB1}

\indent Metric is given by 
\begin{equation}
B_1=u\tilde{l}\tilde{l}+v\tilde{m_1}\tilde{m_1}+w\tilde{m_2}\tilde{m_2} \, ,
\label{b1}
\end{equation}
where $uvw < 0$, not all negative. The coefficient $v$ or $w$ can be rescaled freely. 

There is no general solution for $a$, $b$, and $c$. The scalar curvature is given by 
\begin{equation}
R=-\frac{2[12vw\cos^2\theta+(v-w)^2\sin^2\theta]}{uvw}.
\end{equation}
The matrix $A$ in the equation (\ref{system}) is $4\times3$  and the cases $\theta=0$, $\theta=\frac{\pi}{2}$, and $v=w$ should be considered separately.

\pmb{$\theta=0$}: In this case due to the enlargement of the 
automorphism group, the metric (\ref{b1}) becomes
\begin{equation}
B_1=|z|(\pm\tilde{l}\tilde{l}\pm\tilde{m_1}\tilde{m_1}\pm\tilde{m_2}\tilde{m_2}) \, ,
\end{equation}
which in spacetime coordinates takes the form
\begin{eqnarray} 
ds^2=|z|(\pm d\rho^2+e^{-4\rho}[\pm dx^2+ \pm dy^2]) \, ,
\label{cot2}
\end{eqnarray}
which is either de Sitter for $(-,+,+)$  or AdS for $(+, +, -)$ signs with $R= \pm 24/|z|$ .
Cotton tensor (\ref{cotton}) vanishes identically and we have the relation
\begin{equation}
a=-\frac{4(\pm|z|-c)}{z^2}.
\label{iso(2,0)}
\end{equation}

\pmb{$\theta=\frac{\pi}{2}$}: The coefficients $a$, $b$, $c$ in terms of $u$, $v$, and $w$ are equal to 
\begin{equation}
a=\frac{1}{Q}\cdot \frac{(v-w)^4}{2uvw}, \indent b=\frac{1}{Q}\cdot 4\sqrt{-uvw}(v+w), \indent c=\frac{1}{Q}\cdot 2uvw \, ,
\label{iso(2,pi/2)}
\end{equation}
where $Q= (v-w)^2 + 8(v+w)^2 \neq 0$. In this case the line element is 
\begin{equation}
ds^2=ud\rho^2+(v-w)[\cos(2\rho) dx+\sin(2\rho) dy]^2+w(dx^2+dy^2) \, ,
 \label{iso(2,pi/2) metric}
\end{equation}
which is not familiar to us. Higher order curvature invariants are as follows:
\begin{eqnarray} \nonumber
R_{\mu \nu}R^{\mu \nu}&=&\dfrac{4(v-w)^2(3v^2+2vw+3w^2)}{(uvw)^2} \, ,\\
R_{\mu \rho}R^{\rho \nu}R_{\nu}^{\ \mu}&=&-\dfrac{8(v-w)^6}{(uvw)^3} \, .
\end{eqnarray}
A solution of this type also exists \cite{aliev2, aliev3} in NMG \cite{new}.

%Note that when $b=0$ (i.e., $v=-w$) we get the merger point condition (\ref{merger}).

\pmb{$v=w$}: In this case Cotton tensor (\ref{cotton}) vanishes identically and
\begin{equation}
a=-\frac{4\cos^2\theta(u+c\cdot \cos^2\theta)}{u^2}.
\label{iso(2,v=w)}
\end{equation}
The line element is simply 
\begin{equation}
ds^2=ud\rho^2+ve^{-4\rho \cos \theta}[dx^2+dy^2] \, ,
\label{cot3}
\end{equation}
which is for $u<0$ and $v>0$, either de Sitter if $\theta \neq \pi/2$ or Minkowski if $\theta = \pi/2$.

\subsection{\texorpdfstring{$B_2$-}-type metric}
\label{secB22}

\indent Metric is given by
\begin{equation}
B_2=u\tilde{l}\tilde{l}+\tilde{l}\tilde{m_1}+w\tilde{m_2}\tilde{m_2} \, ,
\label{theta0}
\end{equation}
with $u>0$ and $w \neq 0$.

When $\theta=0$ the metric is Minkowski and the field equation (\ref{MMG}) is solved only if $a=0$. 
For $\theta \neq 0$ the coefficients $a$, $b$, $c$ and 
the scalar curvature $R$ are equal to
\begin{equation}
\label{2}
a=-\frac{2}{9}w\sin^2\theta, \ \ \ \ \ b=\frac{2}{9\sqrt{w}\sin\theta}, \ \ \ \ \ c=-\frac{1}{18w\sin^2\theta}, \ \ \ \ \ R=8w\sin^2\theta.
\end{equation}
The line element is 
\begin{eqnarray} \nonumber
ds^2&=&ud\rho^2+e^{-2\rho \cos\theta}[\cos(2\rho \sin\theta)dxd\rho+\sin(2\rho \sin\theta)dyd\rho] \\
&+&we^{-4\rho \cos\theta}[\sin(2\rho \sin\theta)dx-\cos(2\rho \sin\theta)dy]^2 \, .
\label{2 metric}
\end{eqnarray}
We could not recognize the spacetime it corresponds. Higher order curvature scalars are 
\begin{eqnarray} 
R_{\mu \nu}R^{\mu \nu}=192w^2\sin^4 \theta \, \, , \indent \, \,
R_{\mu \rho}R^{\rho \nu}R_{\nu}^{\ \mu}=512w^3\sin^6\theta \, .
\end{eqnarray}

\section{Solutions on \texorpdfstring{$ISO(1,1;\theta)$}-}

\indent The basis $\{l, m_1, m_2\}$ of $\mathfrak{iso}(1,1;\theta)$ has the brackets
\begin{equation}
[l,m_1]=2\cos\theta m_1+2\sin\theta m_2, \indent [l,m_2]=2\cos\theta m_2+2\sin\theta m_1 \, ,
\end{equation}
and the dual basis is $\{\tilde{l}, \tilde{m_1}, \tilde{m_2}\}$. We choose the group representative as
\begin{equation}
\mathcal{V}=e^{xm_1+ym_2}e^{\rho l} \, ,
\end{equation}
and the Maurer-Cartan one-forms are
\begin{eqnarray} \nonumber
\mathcal{V}^{-1}d\mathcal{V}=(d\rho)l&+&e^{-2\rho \cos\theta}[\cosh(2\rho \sin\theta)dx-\sinh(2\rho \sin\theta)dy]m_1 \\ 
&+&e^{-2\rho \cos\theta}[\cosh(2\rho \sin\theta)dy-\sinh(2\rho \sin\theta)dx]m_2 \, ,
\end{eqnarray}
with $\theta \in [0,\pi/2]$.  When $\theta=0$ the Lie algebras of 
$ISO(1,1;0)$ and $ISO(2;0)$ coincide. Hence,  $\theta=0$ case is already covered in sections \ref{secB1} and \ref{secB22}.
From the automorphism group, two types of metrics can be fixed as below \cite{[Mou]}.

\subsection{\texorpdfstring{$B_1$-}-type metric} \label{iso11}
\indent The $B_1$-type metric is given by 
\begin{equation}
B_1=\delta\tilde{l}\tilde{l}+u(\tilde{m_1}+\tilde{m_2})^2+v(\tilde{m_1}-\tilde{m_2})^2+2w(\tilde{m_1}\tilde{m_1}-\tilde{m_2}\tilde{m_2}) \, ,
\label{b1m}
\end{equation}
with $w^2 > uv$ and $\delta>0$. Two of the parameters $(u,v,w)$ can be set to  $\pm1$ whenever they are non-zero.

\indent The matrix $A$ in (\ref{system}) is $6\times 3$ and there is no general solution for $a$, $b$, $c$. The scalar curvature is
\begin{equation}
R=-\frac{8[3(uv-w^2)\cos^2\theta+uv\sin^2\theta]}{\delta(uv-w^2)}.
\end{equation}
However, the cases $\theta=\frac{\pi}{4}$, $\theta=\frac{\pi}{2}$, and $uv=0$ should be considered separately.

\pmb{$\theta=\frac{\pi}{4}$}: For $w \neq 0$ the coefficients $a$ and $b$ are found as 
\begin{eqnarray} \nonumber
a&=&\frac{2\delta(w^2-uv)(4uv-3w^2)+c(4uv-w^2)(4uv+3w^2)}{3\delta^2(w^2-uv)^2} \, ,\\
b&=&\frac{-\delta(w^2-uv)+c(4uv-w^2)}{3w\sqrt{2\delta(w^2-uv)}}.
\label{iso(1,1,pi/4)}
\end{eqnarray}
The metric  (\ref{b1m}) simplifies to
\begin{equation}
ds^2=\delta d\rho^2+ue^{-4\sqrt{2}\rho}(dx+dy)^2+v(dx-dy)^2+2we^{-2\sqrt{2}\rho}(dx^2-dy^2) \, ,
\label{iso(1,1,pi/4) metric}
\end{equation}
which was identified as timelike or spacelike warped AdS in \cite{[Mou]} depending on signs. 

When $w=0$ the Cotton tensor vanishes identically and we are at the merger point (\ref{merger}) with 
$c=-\delta/4=1/a=4/R$. The metric (\ref{iso(1,1,pi/4) metric}) becomes $(A)dS_2 \times S^1$ that was
found in \cite{[Arv]}, which is clearly not a solution of TMG since $c$ cannot be zero. Its absence
is related to a no-go result on solutions of TMG with a hypersurface orthogonal Killing vector \cite{aliev}.
However, it exists in NMG \cite{new} as was found in \cite{clement2}.

\pmb{$\theta=\frac{\pi}{2}$}: The coefficients $a$, $b$, $c$ in terms of $u$, $v$, and $w$ are equal to
\begin{eqnarray}
a=\frac{2u^2v^2}{\delta(uv+8w^2)(uv-w^2)}, \ \ b=-\frac{2w\sqrt{\delta(w^2-uv)}}{uv+8w^2}, \ \ c=\frac{\delta(uv-w^2)}{2(uv+8w^2)}.
\label{iso(1,1,pi/2)}
\end{eqnarray}

Its spacetime metric is:
\begin{equation}
ds^2 = \frac{\delta}{4} \frac{dr^2}{r^2} +  \frac{d\alpha^2}{r^2} - r^2dt^2 +2wd\alpha dt \, ,
\label{lifshitz}
\end{equation}
where we set $u=1$, $v=-1$ and defined 
\begin{equation}
 r=e^{2\rho} \, , \indent \, \alpha= x+y \, , \indent \, t=x-y.
\label{trans2}
 \end{equation}
Notice that the following constant rescalings $ r \rightarrow \lambda r, \alpha \rightarrow \lambda \alpha,  t \rightarrow \lambda^{-1} t$
leave the metric (\ref{lifshitz}) invariant. When
$w=0$ (which sets $b=0$ and the merger point condition (\ref{merger}) is satisfied) this corresponds to the static Lifshitz 
spacetime with the dynamical exponent $z=-1$  and for $w \neq 0$  it is a stationary 
Lifshitz metric (see \cite{ozgur}). Note that the rotation parameter $w$ is non-zero only when there is a contribution from the Cotton tensor.

\pmb{$u=0$}: The coefficients $a$ and $b$ are  
\begin{equation}
a=-\frac{4\cos^2\theta(\delta +c\cdot \cos^2\theta)}{\delta^2}, \indent b=\frac{w(\delta+2c\cdot \cos^2\theta)}{2\sqrt{w^2\delta}(\cos\theta-2\sin\theta)}.
\label{iso(1,1,u=0)}
\end{equation}
The spacetime metric (\ref{b1m}) takes the form
\begin{equation}
ds^2=\delta d\rho^2+ve^{-4\rho(\cos\theta-\sin\theta)} (dx-dy)^2+2we^{-4\rho\cos\theta}(dx^2-dy^2)  \, ,
\label{iso(1,1,u=0) metric}
\end{equation}
which was identified in \cite{[Mou]} as AdS pp-wave for $\theta \neq \pi/4$ and $\theta \neq \pi/2$. 
When $\theta = \pi/4$ it is AdS and when $\theta = \pi/2$ it is a flat space pp-wave \cite{[Mou]}.

\pmb{$v=0$}: The coefficients $a$ and $b$ are equal to 
\begin{equation}
a=-\frac{4\cos^2\theta(\delta+c\cdot \cos^2\theta)}{\delta^2}, \indent b=-\frac{w(\delta+2c\cdot \cos^2\theta)}{2\sqrt{w^2\delta}(\cos\theta+2\sin\theta)}.
\label{iso(1,1,v=0)}
\end{equation}
The spacetime metric (\ref{b1m}) becomes
\begin{equation}
ds^2=\delta d\rho^2+ue^{-4\rho(\cos\theta+\sin\theta)}(dx+dy)^2+2we^{-4\rho\cos\theta}(dx^2-dy^2) \, ,
\label{iso(1,1,v=0) metric}
\end{equation}
which again corresponds to an AdS pp-wave in general \cite{[Mou]}. But for $\theta = \pi/4$ it is the null warped AdS (Schr\"odinger) spacetime
and when $\theta = \pi/2$ it is a flat space pp-wave \cite{[Mou]}.

%We note that in all the above cases when $b=0$ the merger point condition (\ref{merger}) is satisfied. 

\subsection{\texorpdfstring{$B_2$-}-type metric}

\indent
For $\theta \neq 0$ the metric is given by 
\begin{equation}
B_2=\delta \tilde{l}\tilde{l}+\tilde{l}\tilde{m_1}+u(\tilde{m_1}+\tilde{m_2})^2+v(\tilde{m_1}-\tilde{m_2})^2+2w(\tilde{m_1}\tilde{m_1}-\tilde{m_2}\tilde{m_2}) \, ,
\label{Iso11b2metric}
\end{equation}
with $w^2=uv>0$ and $u+v\neq 2w$. For $w=0$, both Cotton and $J$-tensors vanish and $a=0$ in (\ref{MMG}), which locally corresponds to Minkowski spacetime
as we discuss in section \ref{summary}. 
Hence, we assume $w \neq 0$ which means that $v=w$ is not allowed. One of the coefficients $u$ or $v$ can be set to $\pm 1$.

Using the coordinate transformations given in (\ref{trans2}) the line element becomes
\begin{equation}
ds^2 = \frac{\delta}{4} \frac{dr^2}{r^2} + u r^{-2n} d\alpha^2 + vr^{-2m}dt^2 +2w r^{-(n+m)} d\alpha dt +\frac{1}{4} r^{-(n+1)} d\alpha dr + 
\frac{1}{4} r^{-(m+1)} dt dr  \, ,
\label{generalizedLifshitz}
\end{equation}
where $n=(\cos \theta + \sin \theta)$ and $m=(\cos \theta - \sin \theta)$. Note that the metric is invariant under the scalings
$ r \rightarrow \lambda r, \alpha \rightarrow \lambda^n \alpha,  t \rightarrow \lambda^m t$. Hence, the solution possesses 
a generalized (anisotropic) Lifshitz symmetry. For $\theta=\pi/2$, it becomes a stationary Lifshitz solution with dynamical exponent $z=-1$ 
similar to the one we found above (\ref{lifshitz}).

The coefficients $a$, $b$, $c$ and the scalar curvature $R$ are
\[a=-\frac{32vw^2\sin^2\theta}{9(v-w)^2}, \indent b=-\frac{|v-w|}{18w\sqrt{v}\sin\theta} \, ,\]
\begin{equation}
\label{3}
c=-\frac{(v-w)^2}{288vw^2\sin^2\theta}, \indent R=\frac{128vw^2\sin^2\theta}{(v-w)^2}.
\end{equation}

The only special case 
that should be considered separately is $\theta=\frac{\pi}{4}$.

\pmb{$\theta=\frac{\pi}{4}$}:  The coefficients $a$ and $b$ are given as 
\begin{equation}
a=\frac{32vw^2[(v-w)^2+168c\cdot vw^2]}{3(v-w)^4}, \indent b=-\frac{(v-w)^2-48c\cdot vw^2}{12w|v-w|\sqrt{2v}}.
\label{B2iso(1,1,pi/4)}
\end{equation}
Its metric is (\ref{generalizedLifshitz}) with $m=0$ and $n=\sqrt{2}$  which corresponds to warped flat \cite{[Mou]}.

%Note that for $b=0$ we have the merger point equality (\ref{merger}).

\section{Summary and Discussion}\label{summary}

In this paper we constructed  homogeneous solutions of MMG several of which are new.
We summarize our results in Table \ref{table} where we include only those 
that are non-trivial in the sense that none of the terms in the MMG field equation (\ref{MMG}) vanishes identically. 
In its last column we give classification of our solutions with respect to the Segre-Petrov type of their traceless Einstein tensor
\begin{equation}
P^{a}_{\;\; b} \equiv R^{a}_{\;\; b}-\dfrac{1}{3}R \delta^{a}_{\;\; b} \, ,
\end{equation}
as was proposed in \cite{pope} to which we refer for details. 

From the Table \ref{table} we see that homogeneous solutions of MMG in comparison to TMG can be grouped into three as follows:

\begin{itemize}
 \item Group 1: Solutions which are type N or D in the Segre-Petrov classification can be obtained from TMG solutions with a redefinition of 
 constants as was shown in \cite{bayram}. Corresponding solutions have the same curvature.
 
 For Type D solutions, namely \{(\ref{spacelike}), (\ref{timelike}), (\ref{squashed}), (\ref{ainf metric}), 
 (\ref{iso(1,1,pi/4) metric}), (\ref{B2iso(1,1,pi/4)})\}, we have
 \bea
  a_{\text{MMG}} &=& a_{\text{TMG}} + c \cdot \frac{1}{48} (R+ \frac{4}{9b_{\text{TMG}}^2})(R+ \frac{4}{3b_{\text{TMG}}^2}) \, , \\
  b_{\text{MMG}} &=& b_{\text{TMG}} - c \cdot \frac{b_{\text{TMG}}}{4}(R+ \frac{4}{9b_\text{{TMG}}^2})  \, .
  \eea

  \begin{table}
\begin{center}
\begin{tabular}{|l|l|l|l|l|l|} \hline
\multicolumn{6}{|c|}{\bf{Homogeneous Solutions}} \\ \hline
\bf{Groups} & \bf{Metric} & \bf{MMG} & \bf{TMG} & \bf{Description} & \bf{Type}\\ \hline
\multirow{7}{*}{$SL(2;\mathbb{R})$} & 
\multirow{3}{*}{111-type} & (\ref{111-metric}) & \cmark ($R$=0) & triaxially deformed AdS & \ \  I$_{\mathbb{R}}$\\ \cline{3-6}
& & (\ref{spacelike}) & \cmark & spacelike warped AdS & D\\ \cline{3-6}
 & & (\ref{timelike}) & \cmark & timelike warped AdS & D\\ \cline{2-6}

& \multirow{2}{*}{12-type} & (\ref{12-metric}) & \cmark ($R$=0) & Kundt  & \ \ \ \ II\\ \cline{3-6}
& & (\ref{12-null warped metric}) & \cmark & null warped AdS & N\\ \cline{2-6}

& \multirow{1}{*}{3-type} & (\ref{3-type metric}), chiral pt. & \ \ \ \ \ \xmark & Kundt & \ \ \ \ \ \ III\\ \cline{2-6}

& \multirow{1}{*}{$1z\bar{z}$}-type & (\ref{1zz-metric}) & \cmark($R$=0) & generic & \ \  I$_{\mathbb{C}}$\\ \hline
 
\multirow{2}{*}{$SU(2)$} & & (\ref{su2-general metric}) & \cmark ($R$=0)& triaxially deformed sphere & \ \  I$_{\mathbb{R}}$\\ \cline{3-6}
& & (\ref{squashed}) & \cmark & stretched/squashed sphere & D \\ \hline

\multirow{1}{*}{$A_{\infty}$} & & (\ref{ainf metric}) & \cmark & warped flat & D\\ \hline

\multirow{2}{*}{$A_0$} 
& $B_2$-type & (\ref{ch2 metric}), chiral pt. & \cmark & logarithmic pp-wave & N\\ \cline{2-6} 
& $B_4$-type & (\ref{1 metric}) & \ \ \ \ \ \xmark & generic & \ \ \ \  II\\ \hline

\multirow{2}{*}{$ISO(2;\theta)$} & \multirow{1}{*}{$B_1$-type} 
 & (\ref{iso(2,pi/2) metric}), ($\theta= \pi/2$) & \ \ \ \ \ \xmark & generic & \ \  I$_{\mathbb{R}}$ \\ \cline{2-6}

& \multirow{1}{*}{$B_2$-type} & (\ref{2 metric}), ($\theta\neq 0$)  & \ \ \ \ \ \xmark & generic & \ \ \ \  II\\ \hline

\multirow{6}{*}{$ISO(1,1;\theta)$} & \multirow{4}{*}{$B_1$-type} & (\ref{iso(1,1,pi/4) metric}), ($\theta= \pi/4)$ & \cmark & space/time-like warped AdS & D\\ \cline{3-6}
& & (\ref{lifshitz}), ($\theta= \pi/2$) & \ \ \ \ \ \xmark & stationary Lifshitz & \ \  I$_{\mathbb{R}}$\\ \cline{3-6}
& & (\ref{iso(1,1,u=0) metric}) & \cmark & pp-wave & N\\ \cline{3-6}
& & (\ref{iso(1,1,v=0) metric}) & \cmark &  pp-wave & N\\ \cline{2-6}

& \multirow{2}{*}{$B_2$-type} & (\ref{generalizedLifshitz}), ($\theta \neq 0$) & \ \ \ \ \ \xmark & generalized Lifshitz  & \ \ \ \  II\\ \cline{3-6}
& & (\ref{B2iso(1,1,pi/4)}), ($\theta= \pi/4$) & \cmark & warped flat & D\\ \hline
\end{tabular}
\end{center}
\caption{Comparison of non-trivial homogeneous solutions of MMG and TMG}
\label{table}
\end{table}

For Type N solutions, that is \{(\ref{12-null warped metric}), (\ref{ch2 metric}), (\ref{iso(1,1,u=0) metric}), (\ref{iso(1,1,v=0) metric})\}, we have 
 \bea
  a_{\text{MMG}} &=& a_{\text{TMG}} - c \cdot \frac{R \, a_{\text{TMG}}}{24}  \, , \\
  b_{\text{MMG}} &=& b_{\text{TMG}} - c \cdot \frac{R \, b_{\text{TMG}}}{12}  \, .
  \eea
 
 \item Group 2: Solutions \{(\ref{111-metric}), (\ref{12-metric}), (\ref{1zz-metric}), (\ref{su2-general metric})\} exist in TMG but
only if the cosmological constant vanishes. Hence, they have $R=0$ in TMG. But for MMG, for these solutions the cosmological constant  is 
proportional to the MMG parameter $c$ and therefore $R \neq 0$ is possible.
 
 \item Group 3: Solutions \{(\ref{3-type metric}), (\ref{1 metric}), (\ref{iso(2,pi/2) metric}), (\ref{2 metric}), (\ref{lifshitz}), (\ref{generalizedLifshitz})\} exist 
 only in MMG. 
 
\end{itemize}

It is interesting to note that for all solutions in Group 2 and 3 we have $R^2=16a/c$.
Moreover, three of the solutions in the third group, that is (\ref{1 metric}), (\ref{2 metric}) and (\ref{generalizedLifshitz}), 
appear when $ac= 1/81$ and $9b^2=-8c$. Whether this particular point in the 
parameter space of MMG has any physical significance like chiral (\ref{chiral})
and merger (\ref{merger}) points remains to be seen. Also, more work is 
required to understand spacetimes that we found in (\ref{1 metric}), (\ref{iso(2,pi/2) metric}) and (\ref{2 metric}).

Within the third group, Lifshitz type solutions, that is (\ref{lifshitz}) and (\ref{generalizedLifshitz}), are especially attractive due to 
their possible holographic applications (for a review see \cite{hol1}). Moreover, only few exact Lifshitz solutions which are stationary are known \cite{ozgur}.
In (\ref{lifshitz}) the dynamical exponent is $z=-1$ and rotation is present only when there is a contribution from the Cotton tensor. 
The second one (\ref{generalizedLifshitz})
enjoys a generalized Lifshitz symmetry where each coordinate scales differently. Such solutions are also very rare. 
In four dimensions, one example was found for Einstein gravity coupled to massive vectors in \cite{hol2} and another one in 
Conformal gravity in \cite{pope2}. It would be interesting to study our solutions from the dual CFT perspective. Another 
related issue is to search for Lifshitz black holes \cite{bh}.

In many of the solutions above it is possible to set $b=0$ by choosing other parameters appropriately, after which remarkably one always 
ends up at the merger point (\ref{merger}). Thus, these are solutions of 
of the MMG theory without the Cotton tensor. The fundamental equation of this specific model can be obtained from MMG (\ref{MMG}) by taking the
limit $\mu \rightarrow \infty$, $\gamma \rightarrow \infty$ while keeping $\gamma/\mu^2$ constant which was considered before in \cite{iran} and 
\cite{[chile]}. For this to be consistent, one should still make sure that Bianchi identity (\ref{divergence}) is satisfied, i.e.
\begin{equation}
V^{\mu}=\epsilon^{\mu \rho \sigma}S_{\rho}^{\ \tau}C_{\sigma \tau} =0 \, .
\label{bianchi}
\end{equation}
For our solutions it turns out that $V^{\mu}$ is identically zero except for the following 3 cases:

i) $A_0$ spacetime with $B_1$-type metric ({\ref{a0b1}):
\begin{equation}
V^{\mu}=[\mp \dfrac{2}{vz^3},0,0] \, .
\label{va0}
\end{equation}

ii) $ISO(2;\theta)$ spacetime with $B_1$-type metric ({\ref{b1}):
\begin{equation}
V^{\mu}=[ -\dfrac{64(v-w)^2\cos\theta \sin^2\theta}{u^3vw},0,0] \, .
\end{equation}

iii) $ISO(1,1;\theta)$ spacetime with $B_1$-type metric ({\ref{b1m}):
\begin{equation}
V^{\mu}=[ \dfrac{64uv\sin4\theta \sin\theta}{(w^2-uv)z^3},0,0] \, .
\end{equation}

Recall from the section \ref{a0} that there is no $B_1$ type solution for $A_0$ which is in agreement with $V^{\mu}$ being non-zero  in 
(\ref{va0}). From our analysis in sections \ref{secB1} and \ref{iso11}  one can easily see that, 
the last two vectors become zero precisely at the solutions we found, independent of the value of $b$. 
Moreover, for all our solutions whenever $b=0$ is possible, then the merger point condition (\ref{merger}) is satisfied.
Exceptions appear only when the Cotton tensor identically vanishes. Hence, we reach to the conclusion that:

{\it When the Cotton tensor is absent in the MMG equation (\ref{MMG}), simply transitive 
homogeneous solutions exist only at the merger point (\ref{merger}) provided that they are not 
conformally flat. They satisfy the Bianchi condition (\ref{bianchi}).}

For our solutions for which the Cotton tensor vanishes, it is useful to recall that conformally flat spacetime solutions of MMG are locally maximally 
symmetric away from the merger point (\ref{merger}) which was proven in \cite{alex}. Indeed, (\ref{cot2}) and  (\ref{cot3}) are (A)dS spacetimes which 
are in general away from the merger point but for a specific choice of the parameter $c$ they also exist at the merger point. For (\ref{cot1}) 
cosmological constant 
vanishes ($a=0$) and hence, we conclude that it must be locally Minkowski since the merger point condition, i.e. $ac=1$, is impossible to satisfy. 
This applies also to (\ref{Iso11b2metric}) with $w=0$. On the other hand at the merger point, a conformally flat solution is not necessarily
maximally symmetric. For example, the metric (\ref{iso(1,1,pi/4) metric}) with $w=0$ corresponds to $(A)dS_2 \times S^1$ that was found in \cite{[Arv]} 
and it exists only at the merger point.

We found that two of our solutions given in (\ref{3-type metric}) and (\ref{ch2}) exist at the the chiral point (\ref{chiral}) of the 
parameter space. Only in the latter it is possible to set $b=0$ in which case one ends up at the merger point (\ref{merger}) as we noted 
above.

In this paper we focused on simply transitive homogeneous spacetimes. A natural generalization would be to allow a non-trivial isotropy group 
as was studied in  \cite{ortiz2} for TMG, and in \cite{[Sia]} for NMG. The following metric that was discussed both in \cite{ortiz2} and \cite{[Sia]} 
has 4-dimensional isometry group with no 3-dimensional simply transitive subgroup:
\begin{equation}
 ds^2= -dt^2 + v^2 (d\theta^2 + \sin^2 \theta d\phi^2) \, .
\end{equation}
This is a solution of the MMG field equation (\ref{MMG}) at the merger point (\ref{merger}) with:
\begin{equation}
 a=\frac{1}{c}=\frac{R}{4}=\frac{1}{2v^2} \, .
\end{equation}
The Cotton tensor vanishes identically and its Segre-Petrov type is D.

It would be interesting to repeat our investigation in models closely related to MMG
\cite{setare, bayram2}. Finally, trying to classify all stationary axi-symmetric solutions of MMG as was done for TMG \cite{grumiller}
using a method developed in \cite{clement} would be worth studying. We hope to explore these issues in near future.

\

\noindent {\Large \bf Acknowledgements}

\noindent 
We are grateful to George Moutsopoulos for many useful discussions. We would like to thank \"Ozg\"ur Sar{\i}o\u{g}lu for a critical reading of an earlier
version of this paper and to Marika Taylor for comments about our Lifshitz solutions. JC and NSD are partially supported by 
the Scientific and Technological Research Council of Turkey (T\"ubitak) project 113F034.

\end{document}